\documentclass[aps,prl,reprint,twocolumn]{revtex4-2}
\usepackage{blindtext}
\usepackage{graphicx}
\usepackage{amsmath}
\usepackage[switch]{lineno}

\begin{document}
\title{Reconstructing charged particle track segments with a quantum-enhanced support vector machine}
\author{Philippa Duckett}
\affiliation{Department of Physics and Astronomy, University College London, Gower St, London, WC1E 6BT, UK}
\affiliation{Center for Data Intensive Science and Industry, University College London, Gower St, London, WC1E 6BT, UK}
\author{Gabriel Facini}
\affiliation{Department of Physics and Astronomy, University College London, Gower St, London, WC1E 6BT, UK}
\affiliation{Center for Data Intensive Science and Industry, University College London, Gower St, London, WC1E 6BT, UK}
\author{Marcin Jastrzebski}
\affiliation{Department of Physics and Astronomy, University College London, Gower St, London, WC1E 6BT, UK}
\affiliation{Center for Data Intensive Science and Industry, University College London, Gower St, London, WC1E 6BT, UK}
\author{Sarah Malik}
\affiliation{Department of Physics and Astronomy, University College London, Gower St, London, WC1E 6BT, UK}
\affiliation{Center for Data Intensive Science and Industry, University College London, Gower St, London, WC1E 6BT, UK}
\author{S\'{e}bastien Rettie}
\affiliation{Department of Physics and Astronomy, University College London, Gower St, London, WC1E 6BT, UK}
\affiliation{European Organization for Nuclear Research (CERN), Geneva 1211, Switzerland}
\author{Tim Scanlon}
\affiliation{Department of Physics and Astronomy, University College London, Gower St, London, WC1E 6BT, UK}
\affiliation{Center for Data Intensive Science and Industry, University College London, Gower St, London, WC1E 6BT, UK}

\begin{abstract}
Reconstructing the trajectories of charged particles from the collection of hits they leave in the detectors of collider experiments like those at the Large Hadron Collider (LHC) is a challenging combinatorics problem and computationally intensive. The ten-fold increase in the delivered luminosity at the upgraded High Luminosity LHC will result in a very densely populated detector environment. The time taken by conventional techniques for reconstructing particle tracks scales worse than quadratically with track density. Accurately and efficiently assigning the collection of hits left in the tracking detector to the correct particle will be a computational bottleneck and has motivated studying possible alternative approaches. This paper presents a quantum-enhanced machine learning algorithm that uses a support vector machine (SVM) with a quantum-estimated kernel to classify a set of three hits (triplets) as either belonging to or not belonging to the same particle track. The performance of the algorithm is then compared to a fully classical SVM. The quantum algorithm shows an improvement in accuracy versus the classical algorithm for the innermost layers of the detector that are expected to be important for the initial seeding step of track reconstruction.

\end{abstract}

\maketitle

\section{Introduction}

The Large Hadron Collider (LHC) is currently the highest energy particle collider in the world. It accelerates beams of protons to almost the speed of light and then collides them at a centre-of-mass energy of 13.6 TeV at the centre of large, multi-purpose particle detectors that are designed to reconstruct the outcome of those collisions.  Among the key physics objectives of the LHC are precise measurements of the properties of the Higgs boson, shedding light on the elusive particle(s) that may constitute dark matter, and searching for a wide breadth of new physics phenomena beyond the Standard Model (SM) via exotic decay signatures like long-lived particles. 

To attain these physics goals, the LHC is preparing for an upgrade that will deliver an order of magnitude more data to the experiments by increasing the intensity of the proton beams, resulting in a higher instantaneous luminosity and thus many more collisions taking place every time the proton bunches cross \cite{https://doi.org/10.48550/arxiv.1902.10229}. At this upgraded High Luminosity LHC (HL-LHC) the number of concurrent, overlapping proton-proton interactions (pileup) is expected to reach up to 200, a significant increase from the current average pileup of 40. Such a step change in the running conditions of the collider will significantly increase our capabilities to fulfil the goals of the LHC programme. However, it also presents challenges. The significant increase in detector occupancy will impact the performance of the entire pipeline, including data acquisition, processing, and analysis, as well as simulating the collisions in the detector. This presents significant overhead on the computational resources, with some elements, such as reconstructing charged particle trajectories, becoming a major bottleneck.

To address these high demands on the computational resources, numerous approaches are being pursued, ranging from the development of more efficient algorithms and the application of state-of-the-art machine learning techniques to the use of graphics processing units (GPUs)~\cite{ATLAS:2802799,Contardo:2020886} to execute code that is parallelizable. One of the intriguing new avenues being pursued to tackle these challenges is quantum computing. This new paradigm offers a fundamentally new form of computing by leveraging the phenomena of quantum mechanics and opens the prospect of significantly speeding up our current algorithms and performing calculations that could only be done to some approximation with classical computers. 

Particle physics has seen a surge of interest in ascertaining how quantum computers may impact the future of the field and establishing the scenarios in which they may be most advantageous. The current Noisy Intermediate Scale Quantum (NISQ) devices~\cite{Bharti_2022}, while a stepping stone on the way to universal, fault-tolerant quantum computers, have enabled many of these proof-of-principle studies to be performed. This exploratory phase of applying current NISQ era quantum computers to challenging problems in particle physics will pave the way for the emergence of new ideas and techniques needed to fully exploit quantum computation and identify the specific problems for which they are most suitable. 

Quantum computing algorithms have been studied for a range of different scenarios in high energy physics. The calculation of simple scattering processes via the helicity spinor formalism and the simulation of a parton shower was demonstrated in~\cite{Bepari_2021}. A quantum walk framework was proposed in~\cite{Bepari:2021kwv}, demonstrating that the parton shower is more naturally and efficiently simulated using a quantum walk in two dimensions. Quantum computing has also been applied to jet clustering~\cite{Wei:2019rqy,Pires:2021fka,PhysRevD.106.036021}, classification of collisions of interest from those that are not~\cite{Terashi:2020wfi,dur34540,Wu:2021xsj}, and anomaly detection in searches for new physics~\cite{Ngairangbam:2021yma}.  

The challenging task of connecting the hits left by charged particles in the tracking detector and associating them with the same particle has been studied from several different perspectives including: quantum associative memory to store all the different track patterns and subsequently employ Grover's search algorithm to search through the database and recall the right track pattern~\cite{Shapoval:2019txi}; quantum graph neural networks~\cite{Tuysuz:2020ocw,Tuysuz:2020gjh}; and quantum annealing devices to minimise an objective function~\cite{Zlokapa:2019tkn,refId0,bapst2020pattern}. 

This paper approaches the problem of track reconstruction by proposing a hybrid quantum-classical algorithm that uses a support vector machine with a quantum-estimated kernel. The problem is decomposed into that of classifying short segments of tracks. Often such segments can form the `seeds' for extrapolating to the full trajectory of the track. This `seeding' step is expected to be a large consumer of CPU time at the HL-LHC~\cite{ATLAS:2802799}. Simplifications are implemented to fit the limitations of the presently available quantum simulators.

\section{Data preprocessing}

\begin{figure}[h!]
    \includegraphics[width=0.5\textwidth]{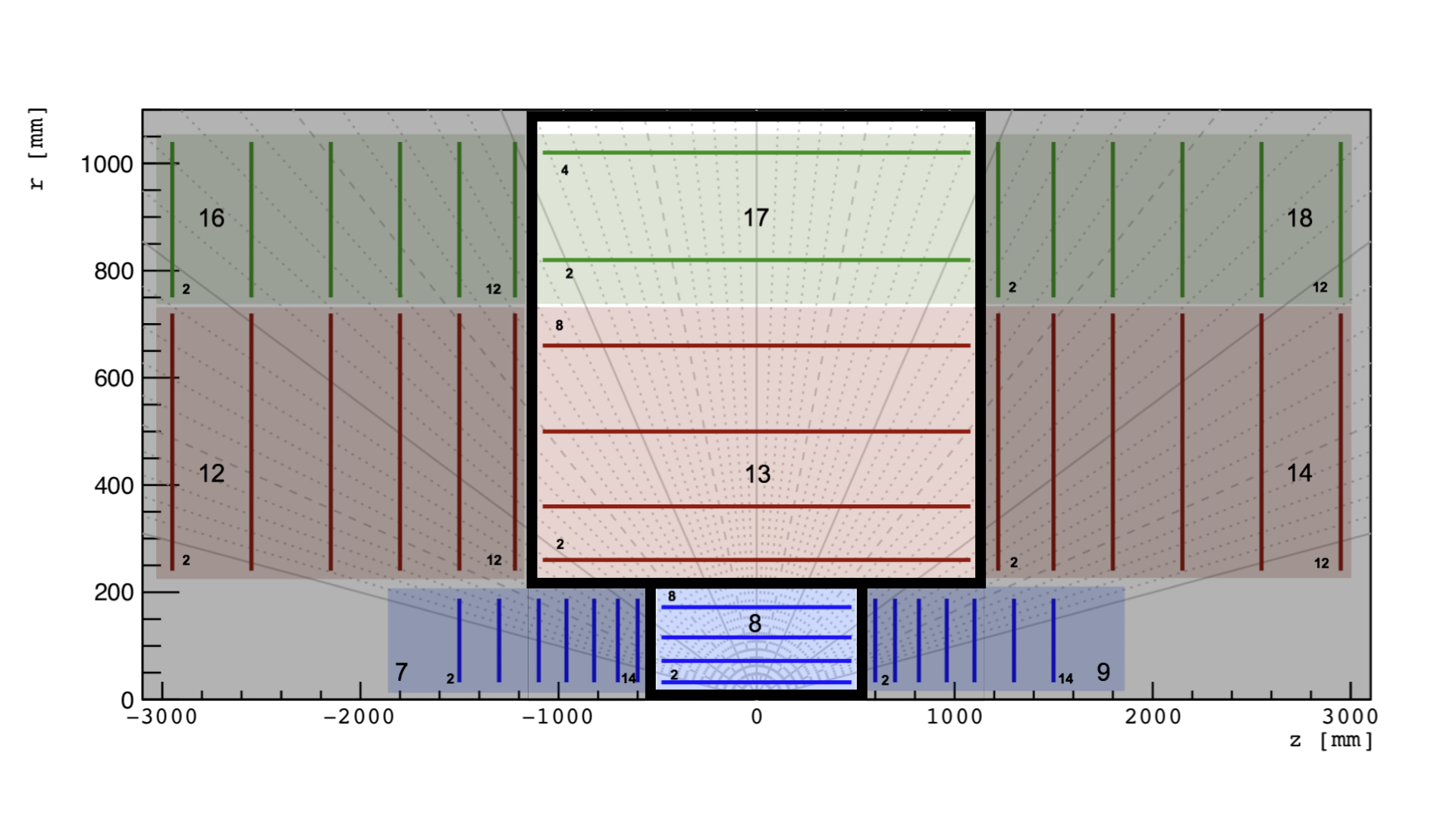}
    \caption{A schematic of the virtual general-purpose detector simulated in the TrackML challenge and the coverage of each sub-detector in the r-z plane. Highlighted is the barrel region used in the analysis. The numbers indicate the various detector components and layers respectively. Original image is taken from~\cite{amrouche:2019wmx}.}
    \label{fig_detector}
\end{figure}

This study utilises the TrackML dataset~\cite{amrouche:2019wmx,amrouche2021tracking} which has been widely used for proof-of-principle studies of classical machine learning algorithms and quantum-based approaches. The dataset provides a simplified simulation of the detector geometry and conditions expected at the HL-LHC. It features a silicon tracking detector with 9 cylindrical layers in the central region and disk geometry in the forward regions, which is typically representative of the ATLAS~\cite{ATLAS:2008xda} and CMS detectors~\cite{CMS:2008xjf}. The detector is segmented into three sub-detectors differing in their spatial resolution, with the inner pixel detector comprising of 4 layers, followed by a short strip detector of 4 layers and then a 2-layer long strip detector. These tracking detectors are immersed in a strong magnetic field aligned with the direction of the proton-proton beam, so charged particles moving through these detectors will typically follow an approximately helical trajectory and show curved trajectories in the transverse $x-y$ plane, the plane perpendicular to the beam line. Figure~\ref{fig_detector} shows the layout of this virtual detector used to produce the TrackML dataset and the coverage of each sub-detector in the $r-z$ plane, where $r$ is the radial dimension and measures the distance from the beam line and $z$ is the distance along the beam line. For the analysis in this paper, only the hits in the barrel region of the detector are used to reduce the total number of hits to a level that can be processed within the current computational constraints. 

The TrackML dataset contains 10,000 simulated events. The process of interest is top-antitop production and overlaid on this `signal' are 200 additional proton-proton collisions to simulate the conditions expected at the HL-LHC. This results in an average of 100,000 hits per event in the tracking detector which must be associated with approximately 10,000 tracks. 

\begin{figure}[h!]
\includegraphics[width=4.0in]{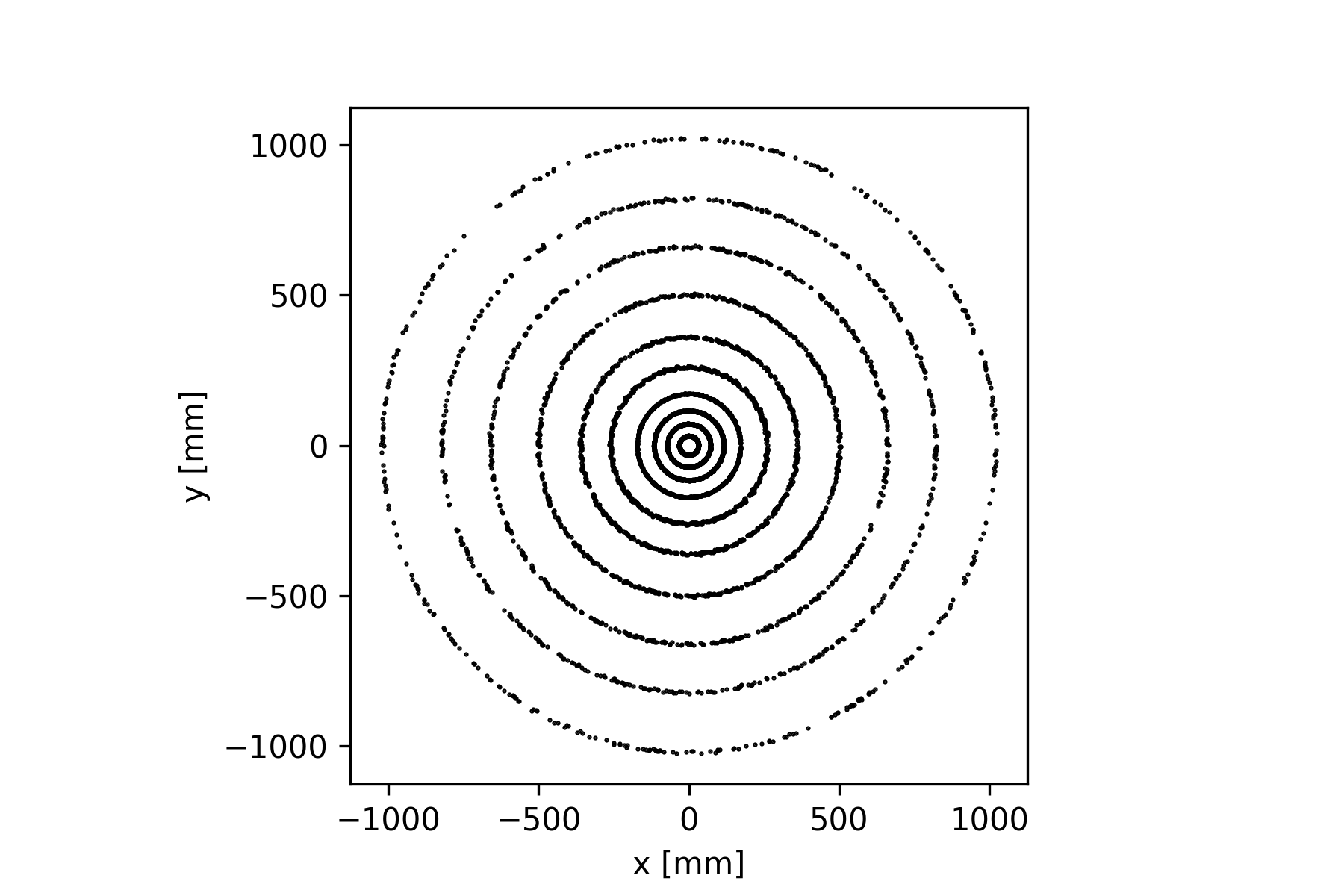}
\includegraphics[width=4.0in]{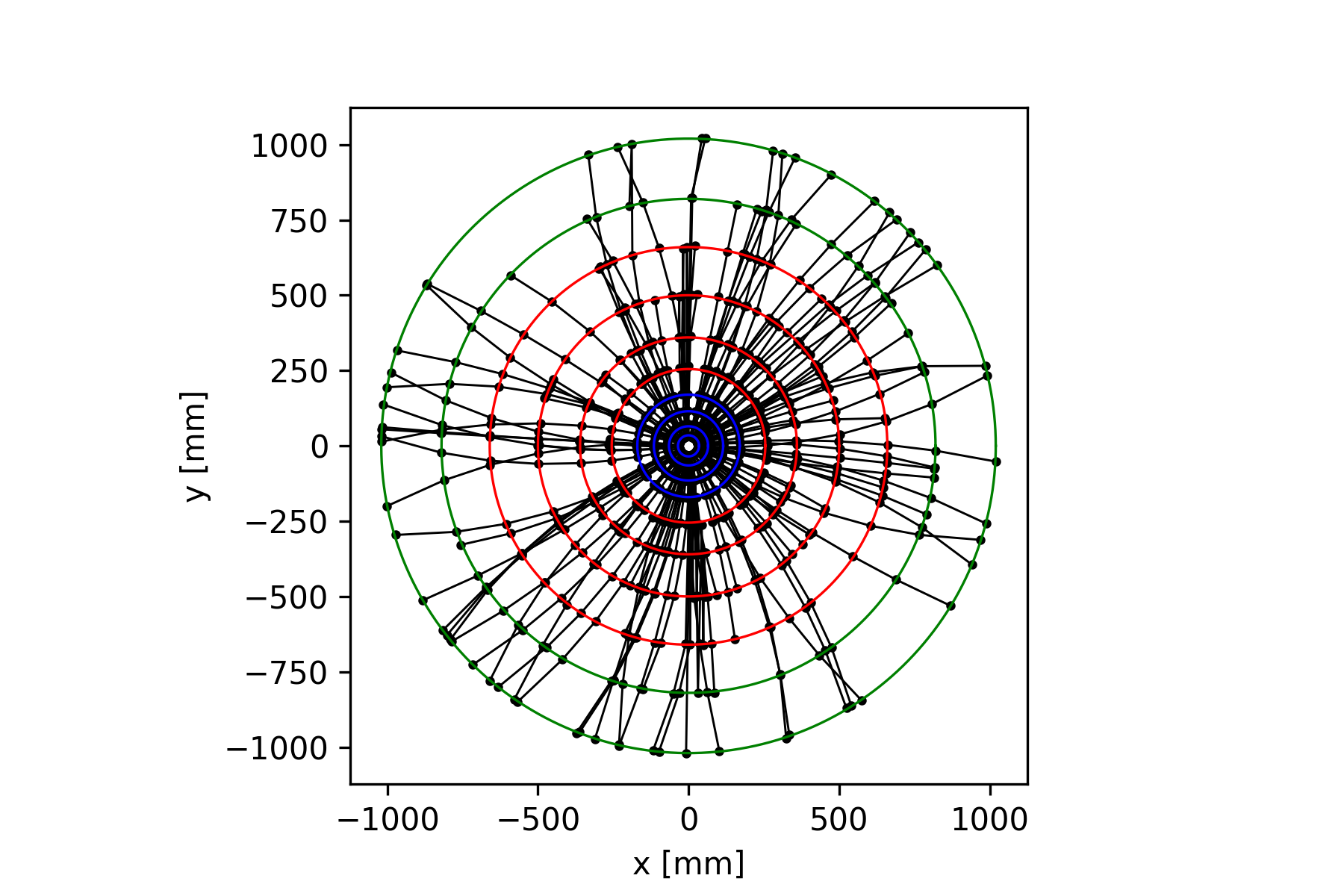}
\caption{The 6518 hits in an example event in the $x$--$y$ detector plane (top) and a fraction of the true tracks reconstructed from those hits (bottom). The hits come from 879 particles which produced triplets in the barrel region. The 10 layers of the detector for the barrel region are also shown. The blue, red, and green layers correspond to the pixel, short strip and long strip detectors, respectively.}
\label{fig_hits}

\end{figure}

The 3-dimensional spatial information for every hit in the detector is provided and this information is used to build the track candidates. The total number of possible combinations of those hits that can lead to a track is very large. Identifying the correct combination of hits that reconstruct the true trajectory of a particle is thus a challenging combinatorics problem. Figure~\ref{fig_hits} illustrates this by showing all hits for an event in the $x$--$y$ plane of the detector and a fraction of true reconstructed tracks formed from a combination of those hits. 
To avoid unphysical hit sequences which would dominate the resulting dataset, selection criteria are applied to reduce the number of possible connections between hits in each event such that they can be processed without overburdening the computational resources. In addition, the problem is formulated as a classification task, with track segments consisting of a set of 3 hits in adjacent layers of the detector, called triplets, being classified as belonging to a single particle track or not. 

The hits are described by three coordinates; $r,\phi$, and $z$, where $\phi$ is defined as the angle around the $z$ axis. A total of 300 events have been processed for classification. The first step in constructing the triplets is to make a dataset of doublets, which are defined as two consecutive hits in the detector. Selection criteria are applied to reduce the size of the doublet dataset and improve its quality. The following observables are used in the selection; the intercept from the extrapolation of the doublet to the $z$ axis, $z_{0}$, and the ratio $\frac{\Delta \phi}{\Delta r}$, as calculated from the difference in $\phi$ and $r$ between each hit forming the doublet. This selection is summarised in Table~\ref{tab:edge_cuts}, and was originally implemented in \cite{Farrell:2018cjr}. 

The selection of triplets is based on the estimation of the transverse momentum ($p_{T}$) as determined from the three hits, the $\theta$-breaking angle and the $\phi$-breaking angle. The angle $\theta$ is defined in the $r$-$z$ plane and a breaking angle is that between the straight lines (connecting the two hits in a doublet) of two doublets that form a triplet. The triplet selection is summarised in Table~\ref{tab:triplet_cuts}. 

\begin{table}[h!]
    \begin{tabular}{c|c} 
     Variable & Selection\\ \hline
    $\frac{\Delta \phi}{\Delta r}$ & $\le 6 \times 10^4 $ [$\frac{\text{rad}}{\text{mm}}$] \\
    $|z_0|$ & $\le$ 100 [mm] \\
    \end{tabular} 
    \caption{The selection criteria applied to select doublets, using the $z_{0}$ intercept from the extrapolation of the doublet to the $z$ axis and the ratio of the difference in $\phi$ and $r$ between each hit forming the doublet.}
    \label{tab:edge_cuts}
    \end{table}

\begin{table}[h!]
\begin{tabular}{c|c}
Variable  & Selection\\ \hline
$\theta$-breaking angle & $\le 0.05-0.07$ [rad] \\
$\phi$-breaking angle & $\le 0.05 - 0.12$ [rad]\\
$p_T$ & $\ge$ 0.75 [GeV] \\
\end{tabular} 
\caption{The selection criteria applied to select triplets based on the estimated $p_{T}$ and the $\theta$ and $\phi$ angles between two doublets that form a triplet. A range of values is given when the selection depends upon detector components traversed.}
\label{tab:triplet_cuts}
\end{table}

\section{Support vector machine}
\begin{figure}[h!]
    \includegraphics[width=3.5in]{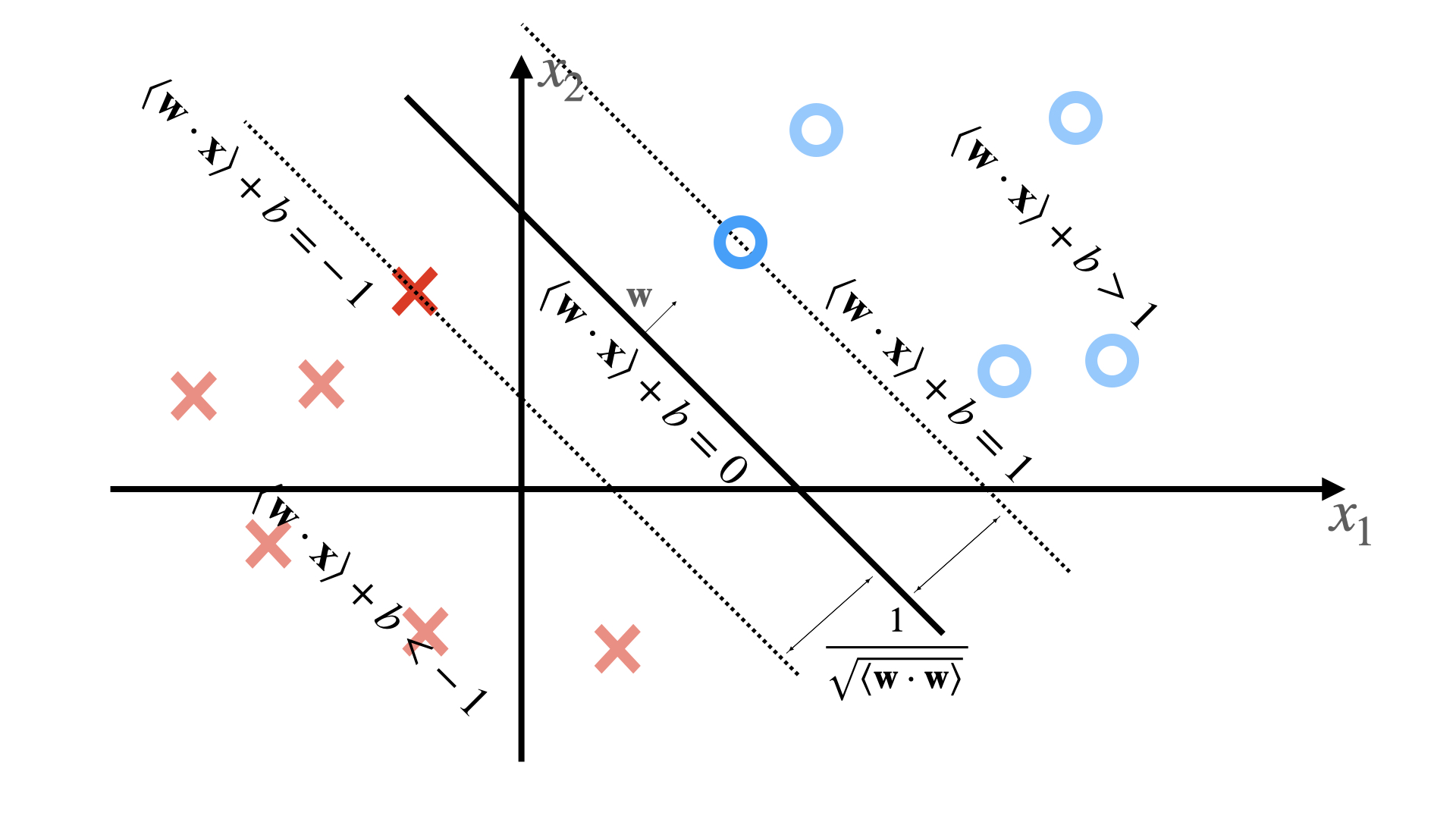}
    \caption{A visual representation of two classes of data in a 2-dimensional space, separated by a hyperplane $\langle \mathbf{w}\cdot\mathbf{x}\rangle+b$ (solid line). The highlighted points lying closest to the separation plane are called support vectors and the dotted lines passing through them define the margins.}
    \label{fig:svm}
\end{figure}

The proposed algorithm utilises a support vector machine (SVM) \cite{cristianini_shawe-taylor_2000}, where a kernel function is calculated either on a (simulated) quantum or a classical computer. A support vector machine is a supervised machine learning algorithm that classifies data by drawing linear decision boundaries (hyperplanes) between different groups of data. This paper focuses on discriminating between two classes of data. It takes a training dataset of size $N$ of the form $(\mathbf{x}^{1},y^{1}),\ldots ,(\mathbf{x}^{N},y^{N})$, where $\mathbf{x}^{i}$ is an $M$-dimensional vector and $y^{i} =\pm 1$ for data that belongs to one of two classes. The hyperplane is defined by $\langle \mathbf{w} \cdot \mathbf{x} \rangle + b = 0$, where $\mathbf{w}$ is the normal vector to the hyperplane and $b$ is an offset. These parameters are determined during the learning process. For the simple case of linearly-separable data, the training points $\mathbf{x}^{i}$ of the two classes are placed on either side of the decision boundary, satisfying $f(\mathbf{x}^{i}) = \mathrm{sign}\left(\langle \mathbf{w} \cdot \mathbf{x}^{i} \rangle + b\right) = y_i$, where $f(\mathbf{x})$ is called the decision function.
The points closest to the hyperplane are called support vectors and the distance between them and the hyperplane is called the margin. The goal is to optimise the parameters of the hyperplane such that the margin is maximised. Figure~\ref{fig:svm} shows a visual representation of this. Once the hyperplane has been found, a previously unseen data point $\mathbf{z}$ can be classified using the decision function.

\begin{figure}[h!]
    \includegraphics[width = 3.5in]{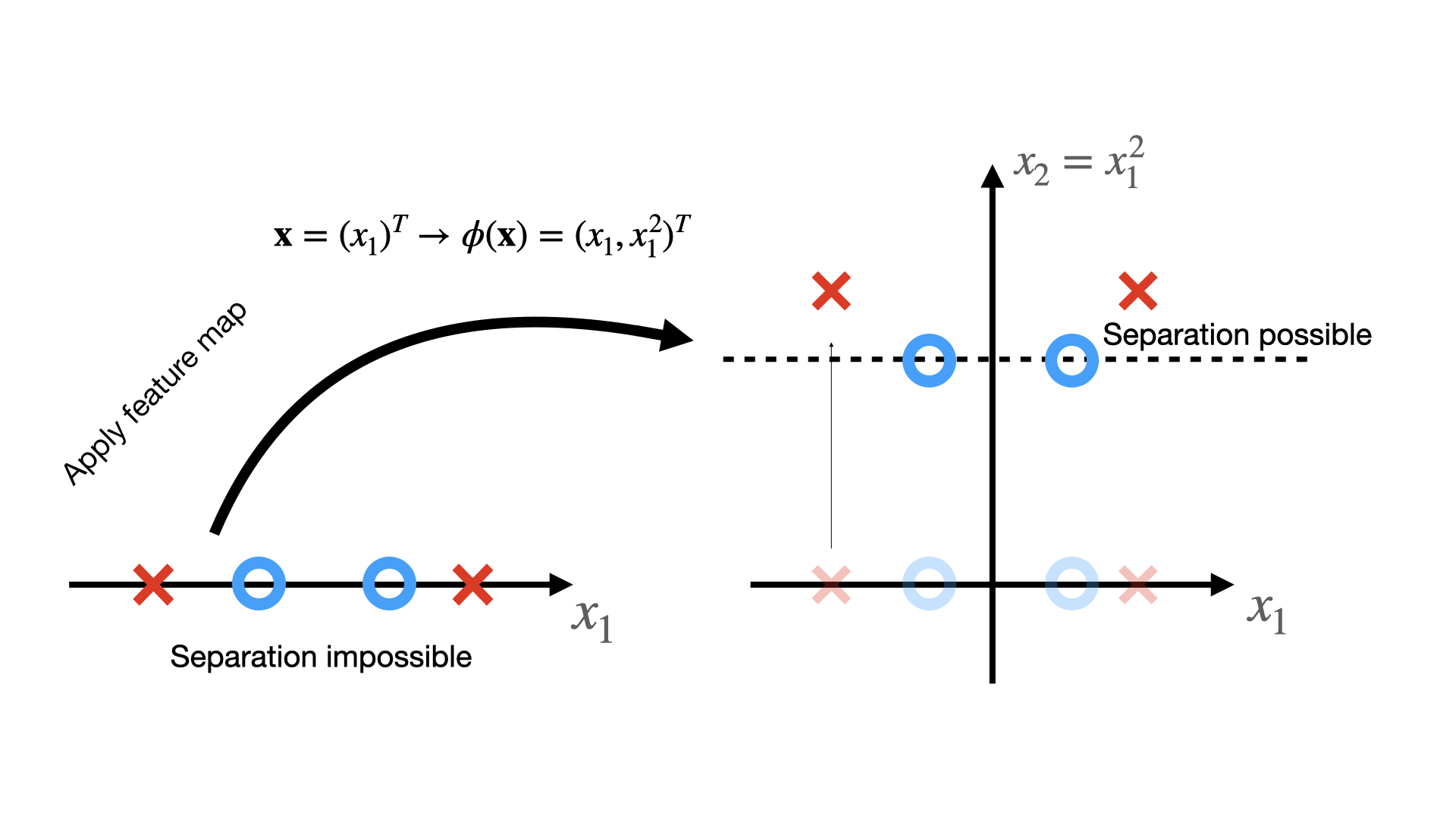}
    \caption{A visual representation of a simple feature map that takes an inseparable dataset in one-dimension to a two-dimensional feature space. Separation with a linear hyperplane is possible in the new feature space.}
    \label{fig:feature_map}
\end{figure}

The decision boundary is usually defined not in the original data space but in a higher-dimensional \textit{feature space} obtained with a feature map $\phi(\mathbf{x})$. This can introduce non-linearity whilst keeping the decision boundary linear. The goal of this operation is to achieve better separation of the two classes. Figure~\ref{fig:feature_map} shows a simple example of a feature map's functionality. 
SVMs are an example of a kernel method, where the kernel $k(\mathbf{x}, \mathbf{y}) = \langle \phi(\mathbf{x}) \cdot \phi(\mathbf{y}) \rangle$ is a function with arguments in the original space of the data, defining a distance measure between two points in the feature space. The remarkable property of this function is that it returns the inner product in the feature space, sidestepping the explicit application of the feature map, which can become computationally expensive for sophisticated feature spaces. In support vector machines, this property can be utilised to find the separation hyperplane. This is possible because linear learning machines can be expressed in a dual representation, following the Karush-Kuhn-Tucker theory \cite{sundaram_1996}. During the optimisation of the dual problem, one needs to find a kernel matrix $K_{\mathbf{x,y}} = k(\mathbf{x},\mathbf{y})$ (an $N \times N$ symmetric matrix) from all pairs of training data points. 
Expressed in its dual form, the decision function becomes:
\begin{equation}
    y(\mathbf{x}) =\mathrm{sign}\left(\sum_{i=1}^{N}y^{i}\alpha^{i}k(\mathbf{x^{i}},\mathbf{x}) + b\right),
    \end{equation}
where $\alpha^{i}$ are the coefficients which need to be optimised.
Just like quantum computing, kernel methods perform implicit computations in a possibly intractably-large Hilbert space through the efficient manipulation of data inputs.

\section{Quantum kernel estimation}
Quantum computers can be utilised in kernel methods if one considers a quantum circuit $\mathcal{U}(\mathbf{x})$ whose gates are parametrised by the original features of some classical data. The result of such a circuit before measurement is a quantum state which exists in a higher-dimensional Hilbert space. 
This is equivalent to a feature map. The quantum state is defined as~\cite{Schuld2021}:
\begin{equation}
    \mathbf{x} \rightarrow \rho(\mathbf{x}) = |\psi(\mathbf{x})\rangle \langle \psi(\mathbf{x})|,
\end{equation}
where $| \cdot \rangle$ denotes the usual Dirac vector and $\rho(\mathbf{x})$ is obtained via
\begin{equation}
    \rho(\mathbf{x}) = \mathcal{U}(\mathbf{x})\rho_{0}\mathcal{U}^{\dagger}(\mathbf{x}),
\end{equation}
with an initial state $\rho_0$. An all-zero initial state is used with $|\psi_0\rangle = |0^{\otimes M}\rangle$.
The kernel associated with such a feature map is obtained from~\cite{Havl_ek_2019}:
\begin{equation}
    k(\mathbf{z}, \mathbf{x}) = \text{tr}\{\rho(\mathbf{z}) \rho(\mathbf{x})\} = \left|\langle\psi(\mathbf{z})|\psi(\mathbf{x})\rangle\right|^2.
    \label{eq:quantum_kernel}
\end{equation}
This inner product can be calculated from the transition amplitude of two states;
\begin{equation}
    \left|\langle\psi(\mathbf{z})|\psi(\mathbf{x})\rangle\right|^2 = \left| \langle0^{\otimes M}|\mathcal{U}^{\dagger}(\mathbf{z})\mathcal{U}(\mathbf{x})|0^{\otimes M} \rangle  \right|^2.
\end{equation}
The circuit $\mathcal{U}^{\dagger}(\mathbf{z})\mathcal{U}(\mathbf{x})|0^{\otimes M}\rangle$ is run repeatedly over $R$ identical runs (shots). The fraction of measurements yielding an all-zero output gives an estimation of the kernel function for the two points $\mathbf{x}$ and $\mathbf{z}$, which forms an entry in the kernel matrix. Repeated evaluations for all combinations of the input dataset give the full kernel matrix.
Similar states have large kernel matrix entries while orthogonal points give $k(\mathbf{x}, \mathbf{z}) = 0$.

Feature maps of particular interest are those that are difficult to calculate using classical means whilst providing good classification of data. Ideally, a kernel matrix resulting from Eq. \ref{eq:quantum_kernel} would produce results better than any classical classifier and be calculated significantly faster on a quantum device. The kernel function proposed in~\cite{Havl_ek_2019} is based on the 3-fold forrelation (`Fourier correlation') problem~\cite{doi:10.1137/15M1050902}. The function is conjectured to have an exponential separation in complexity between its quantum and classical estimation. Further discussion of the potential for speedup is presented later.

\begin{figure}[!h]
    \includegraphics[width=3.6in]{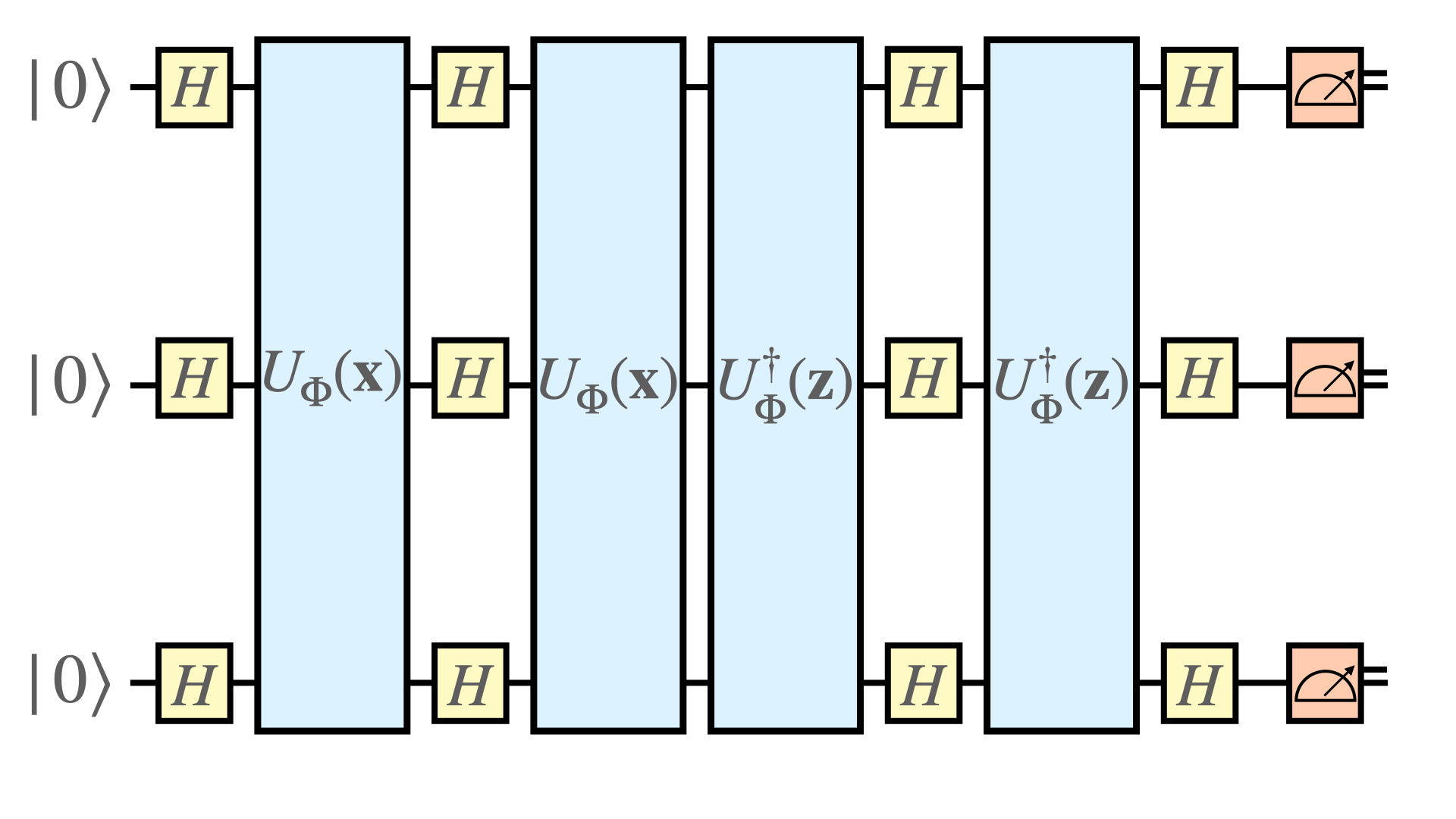}
    \caption{Quantum circuit diagram used to estimate the kernel and determine the inner product between two quantum states shown for data with three features. }
    \label{fig:circuit1}
    \end{figure}

\begin{figure}[!h]
    \includegraphics[width=3.2in]{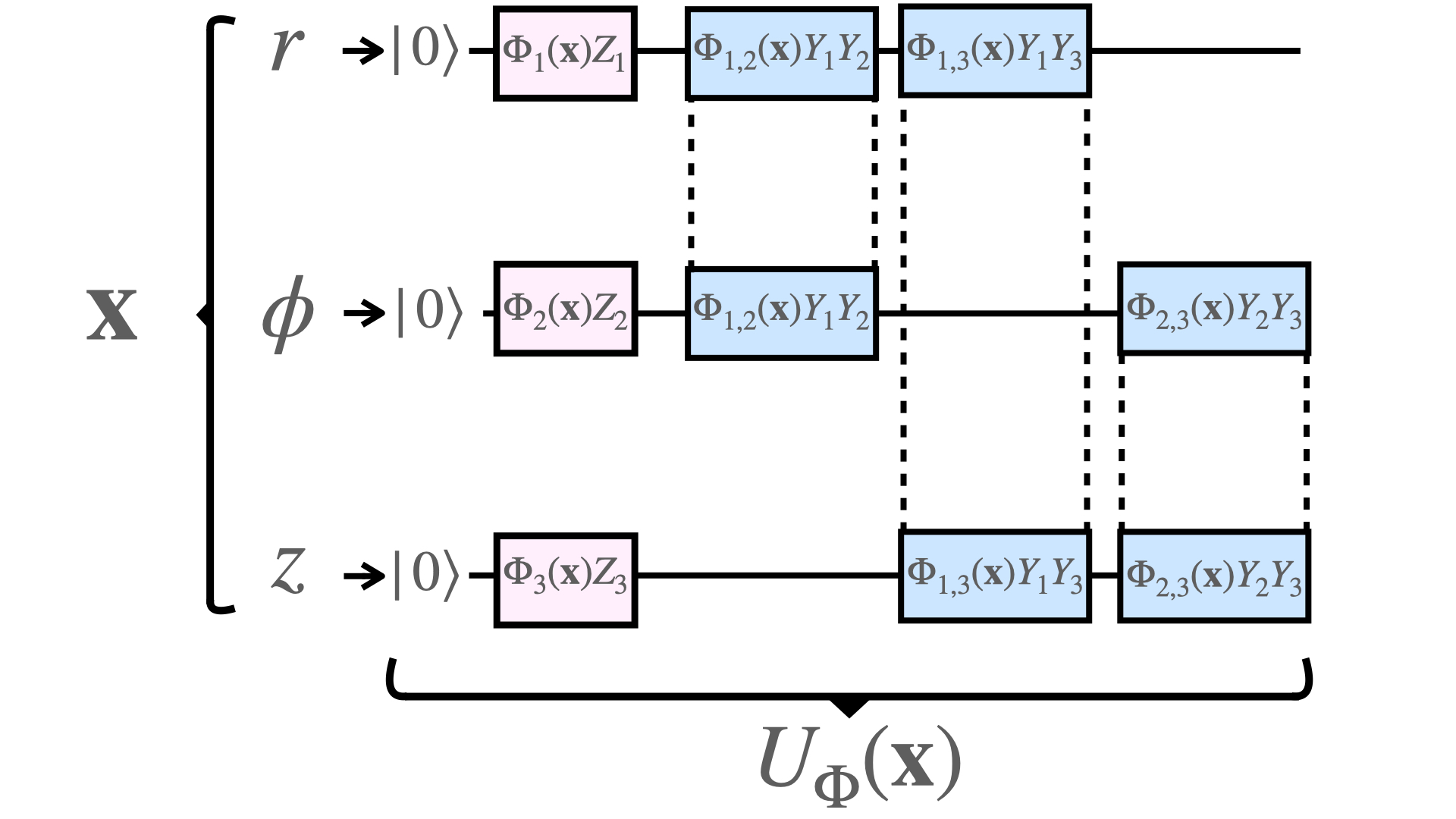}
    \caption{Circuit diagram used to calculate $U_{\phi(\mathbf{x})}$ in the full circuit, shown for a datapoint with three features, which correspond to the spatial coordinates of a single hit in the tracking detector. The single-qubit gates are shown in pink and two-qubit gates in blue.}
    \label{fig:circuit2}
    \end{figure}

The kernel-generating circuit is of the form $\mathcal{U}(\mathbf{x}) = U_{\phi(\mathbf{x})} H^{\otimes M} U_{\phi(\mathbf{x})} H^{\otimes M}$ where $H$ is the Hadamard gate and 
\begin{equation}
    U_{\phi(\mathbf{x})} = \exp{
        \left( 
        i\sum_{S \subseteq [M]}{\phi_S(\mathbf{x})\prod_{i \in S}{Z_i}}
        \right)}.
    \label{eq:havlicek_map}
\end{equation}
$Z_i$ is a gate rotating the $i$-th qubit around the $Z$ axis on the Bloch sphere by an amount defined by $\phi_S(\mathbf{x})$.
$S$ denotes a subset of qubits. Only subsets with $|S|\leq 2$ are considered. The circuit for kernel estimation with $\mathcal{U}(\mathbf{x})$ in the case of 3-dimensional data is shown in Fig.~\ref{fig:circuit1}.
Ideas for generalising the circuit have been proposed in \cite{Shaydulin_2022} and \cite{https://doi.org/10.48550/arxiv.2012.07725}. Following the latter, we implement unitaries of the form:
\begin{equation}
    U_{\phi(\mathbf{x})} = \exp{
    \left( 
    i\alpha\sum_{S \subseteq [M]}{\phi_S(\mathbf{x})\prod_{i \in S}{\sigma^a_i}}
    \right)},
    \label{eq:upgraded_map}
\end{equation}
where $\sigma^{a} \in X,Y,Z$ and $\alpha$ is a constant factor to regulate the degree of rotation of the qubits. An example of $U_{\phi(\mathbf{x})}$ for 3-dimensional data can be found in Fig.~\ref{fig:circuit2}.

This quantum-estimated kernel is then used as input to a classical support vector machine which performs the training and classification. The full circuit thus takes events of dimension $M$ and projects them into an $2^{M}$-dimensional quantum space where the hyperplane separating the two classes of data is calculated. 

\section{Results}
Results from the quantum-enhanced algorithm presented in this section were obtained with $\alpha = 0.1$, $\phi_{k}(\mathbf{x}) = x_k$, $\sigma^a_k = Z$ for single qubit rotations and $\phi_{l,m}(\mathbf{x}) = (\pi - x_l)(\pi - x_m)$, $\sigma^a_{l,m} = Y_lY_m$ for two-qubit rotations. These were compared to an RBF kernel~\cite{Musavi1992OnTT}, defined as $K_{RBF}(\mathbf{x}, \mathbf{z}) = \exp{\left(-\gamma\lVert \mathbf{x} - \mathbf{z} \rVert^2\right)}$ with $\gamma = 1$. To reduce generalisation error and model complexity, a regularisation term $C$ can be included into the optimisation loss function. The optimal coefficients and classical kernel type were found using a grid search~\cite{scikit-learn} with cross validation and a parameter scan optimising for validation score and training time.
Large values of $C$ result in a larger penalty for overfitting. The optimal value of $C = 10^6$ determined from this optimisation procedure was used in both the classical and quantum kernels. 

The metrics used to quantify the performance of the classifiers are defined below, through the confusion matrix shown in Table ~\ref{tab:confusion}.
\begin{table}[h!]
    \centering
    \begin{tabular}{c|c|c}
         & Predicted positive & Predicted negative \\ \hline 
         Actual positive & True positive (TP) & False negative (FN) \\
         Actual negative & False positive (FP) & True negative (TN) \\
    \end{tabular}
    \caption{Confusion matrix used in defining the performance metrics for the classifiers used in triplet recognition.}
    \label{tab:confusion}
\end{table}
\begin{equation}
    \text{Accuracy} = \frac{\text{TP+TN}}{\text{TP+FP+TN+FN}},
\end{equation}
\begin{equation}
    \text{Efficiency} = \frac{\text{TP}}{\text{TP+FN}},
\end{equation}
\begin{equation}
    \text{Purity} = \frac{\text{TP}}{\text{TP+FP}}.
\end{equation}
A good model is expected to score high in all three metrics. Accuracy gives an overall percentage of correct guesses, efficiency is the fraction of actual true objects correctly recognised and purity measures how often the model mistakes a fake object for a true one.

\subsection{Full detector triplets}

The dataset used in the classification consists of triplets that passed the preprocessing step described above, which ensures a sample purity of 52\% and 80\% for doublets and triplets, respectively and a 99\% sample efficiency in both datasets. An average of 4,600 triplets remain per event. 
The spatial coordinates of each hit in the triplet are used as the input data to the hybrid algorithm. This results in a 9-qubit circuit. To accommodate the dataset into our current computational constraints, it is further divided into 16 equal sections in the $\phi$ plane, with each section subtending $\frac{2\pi}{16}$ radians in $\phi$. A support vector machine is then defined for each of these regions and the relevant quantum kernel estimated. 
The data is divided into 50 events for training and 15 events for testing, equivalent to a total of around 230,000 and 70,000 triplets, respectively. 
The performance of both the classical algorithm and the quantum algorithm are evaluated using the three metrics defined above; accuracy, efficiency and purity. Furthermore, since the preprocessing step selects triplets that are more likely to form track candidates, a benchmark scenario is introduced to illustrate the performance of the classical and quantum algorithms on top of the preprocessed data. Triplets are randomly selected from the dataset and the classical and quantum classifiers are compared against this benchmark to demonstrate the improvements in classification accuracy.

\begin{figure}[!h]
\includegraphics[scale=0.55]{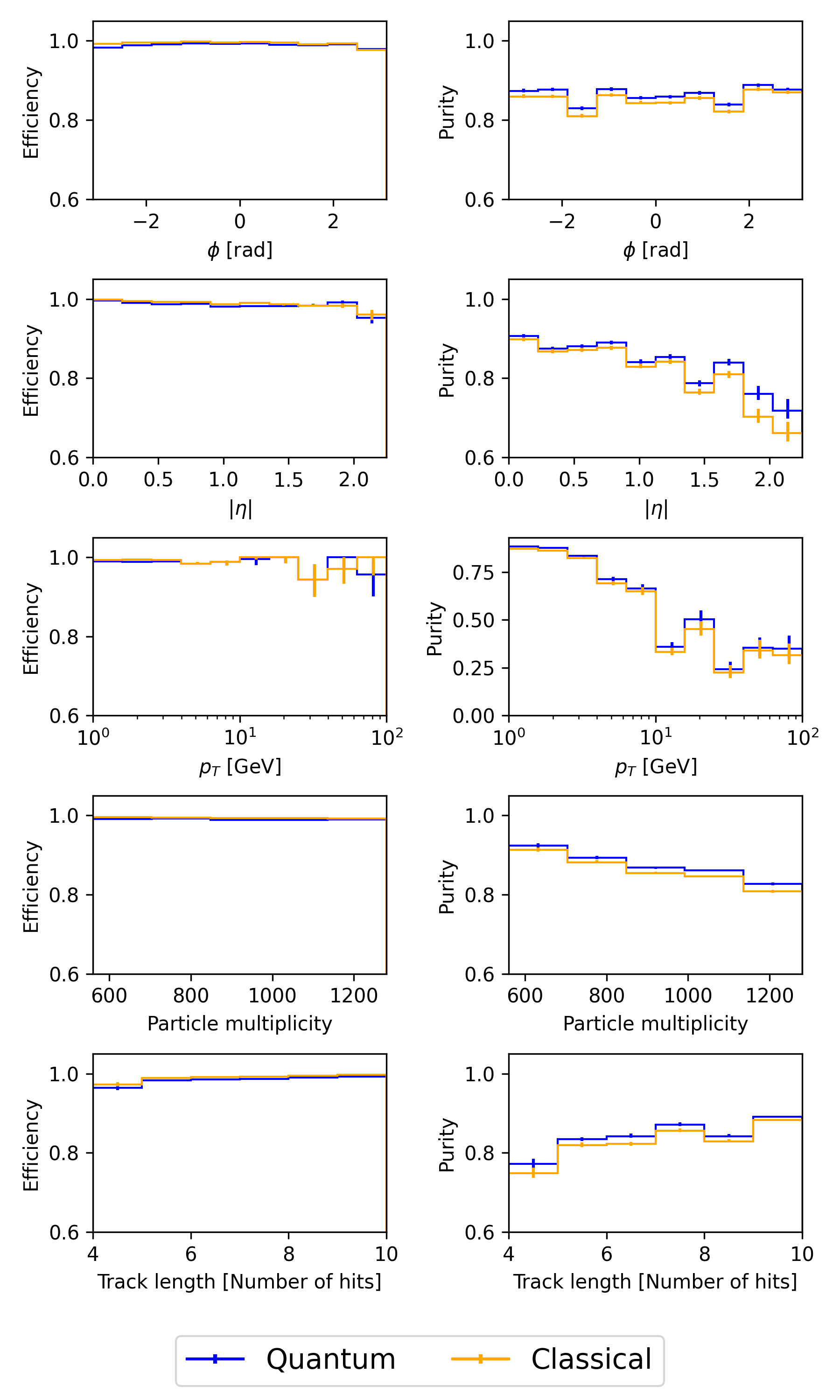}
\caption{Track reconstruction efficiency and purity for triplets in the barrel detector as a function of $\phi$, $|\eta|$, $p_{T}$, particle multiplicity, and the number of hits associated with the track (track length). These are compared for the quantum-estimated kernel and the classical kernel.}
\label{fig:eff_pur_all}
\end{figure}

\begin{figure}[!h]
    \includegraphics[scale=0.26]{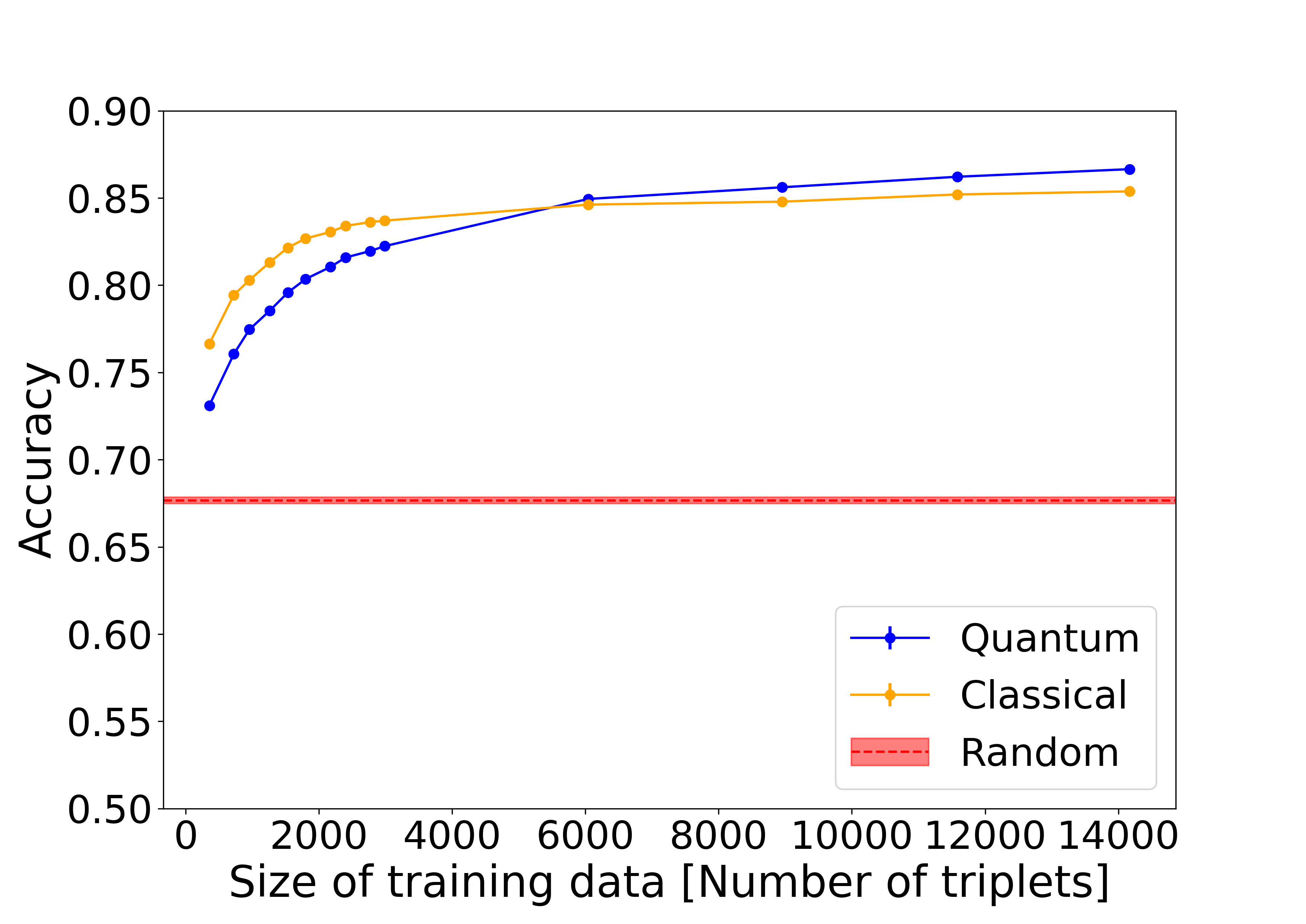}
    \caption{Accuracy to identify triplets in the barrel detector as a function of the size of the training dataset for the quantum-estimated kernel, classical kernel and selecting random triplets from the preprocessed dataset. }
    \label{fig:acc_all}
\end{figure}

Figure~\ref{fig:eff_pur_all} shows the dependence of the efficiency and purity scores on relevant geometric and kinematic properties of the triplets; the angle $\phi$ of the first (innermost) hit of the triplet, the pseudorapidity $|\eta|$ of the triplet, the true $p_{T}$ of the triplet, the number of true particles in the event (particle multiplicity), and the number of hits corresponding to a true track (track length). For fake triplets, there is some ambiguity in determining the true $p_{T}$ and track length, as the three hits can come from three separate particles. In such cases, the choice was made to define these variables using the particle associated with the first hit.
The two classifiers show mostly comparable performance and a similar dependence on the observables defined above. Efficiencies close to 1.0 are achieved for most bins. Reduced values of the purity are observed in regions with reduced number of triplets for training, such as high-$\eta$ and high-$p_T$. 
The accuracy scores of the classical and quantum algorithms as a function of the size of the training data are shown in Fig.~\ref{fig:acc_all}. The training size grows up to the computational limit imposed by the quantum simulator. 
Whilst the overall performance of the two algorithms follows a similar trend, the classical algorithm performs slightly better at low training size and the quantum algorithm shows a small advantage for training size above 6,000 triplets. Both algorithms significantly outperform the benchmark scenario of randomly selecting triplets. 

\begin{figure}[!h]
    \includegraphics[scale=0.27]{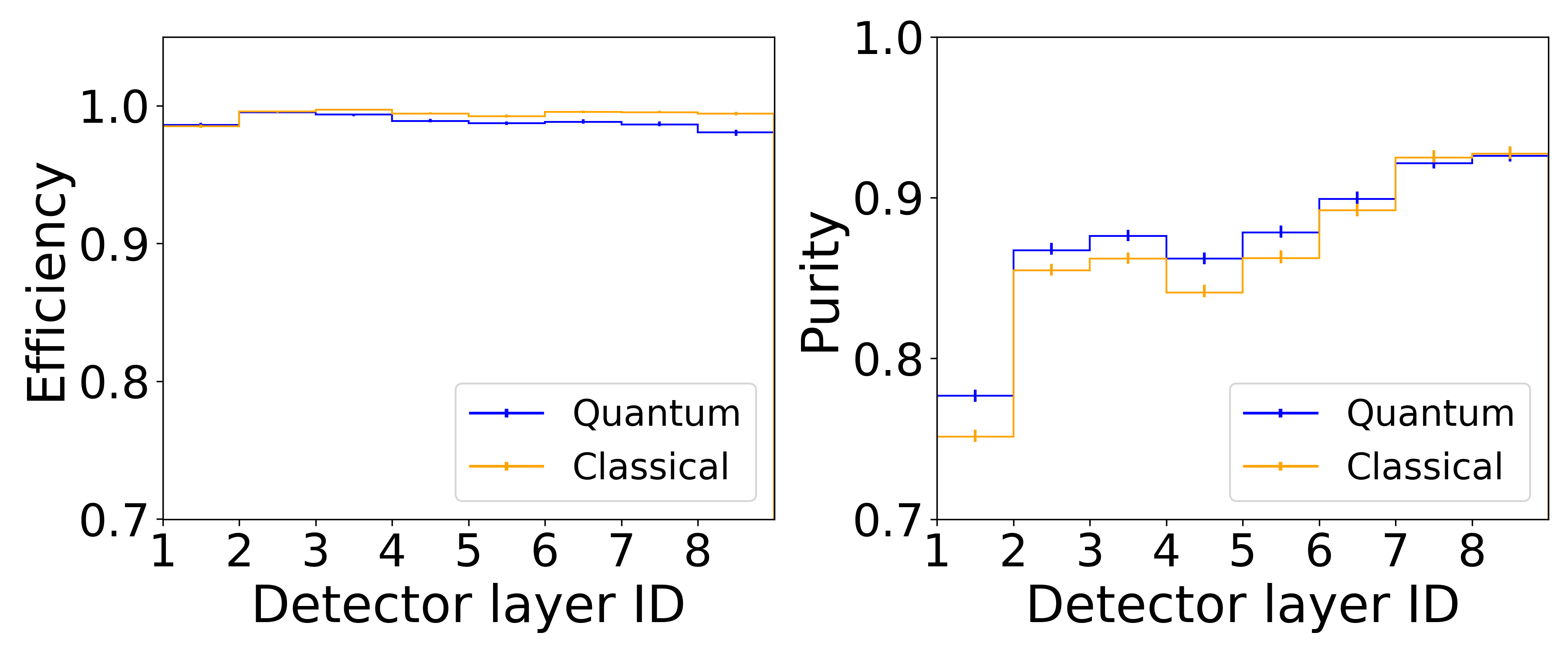}
    \caption{Efficiency and purity as a function of the detector layer, starting from the innermost layer nearest the centre of the detector. Layers 1$-$4 correspond to the pixel detector, layers 5$-$8 belong to the short strip detector and layers 9 and 10 are the long strip detector. Triplets are binned by their innermost hit.}
    \label{fig:eff_pur_layers}
\end{figure}

It is also instructive to study the performance of these algorithms for different layers of the detector, as the detector occupancy progressively decreases from the inner to the outer layers. Figure~\ref{fig:eff_pur_layers} shows the comparison between the quantum and classical algorithms for the efficiency and purity in layers 1-9 of the detector. While the efficiency and purity are similar between the two algorithms for the full detector, the largest difference in purity occurs for triplets identified in the first layer of the detector (the first hit is in the first layer). Since triplet formation is part of the seeding step used in many track reconstruction algorithms, we study the performance of our algorithms for triplets identified in the inner detector, with the first hit being in the first layer of the detector. 

\subsection{Innermost triplets}

\begin{figure}[!h]
    \includegraphics[scale=0.55]{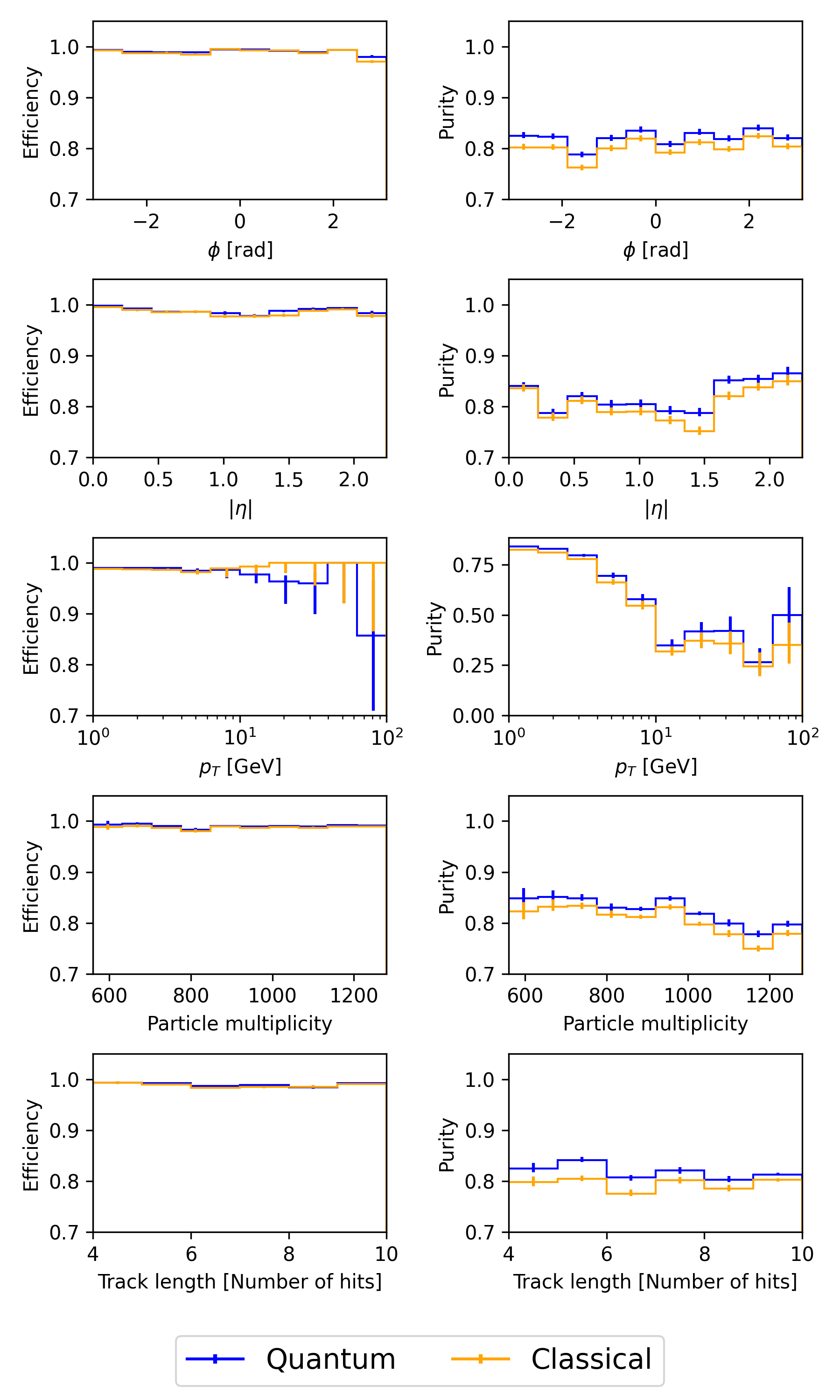}
    \caption{Track reconstruction efficiency and purity for triplets in the inner detector barrel region as a function of $\phi$, $|\eta|$, $p_{T}$, particle multiplicity, and the number of hits associated with the track (track length). These are compared for the quantum-estimated kernel and a classical kernel.}
    \label{fig:eff_pur_inner}
    \end{figure}

\begin{figure}[!h]
    \includegraphics[scale=0.26]{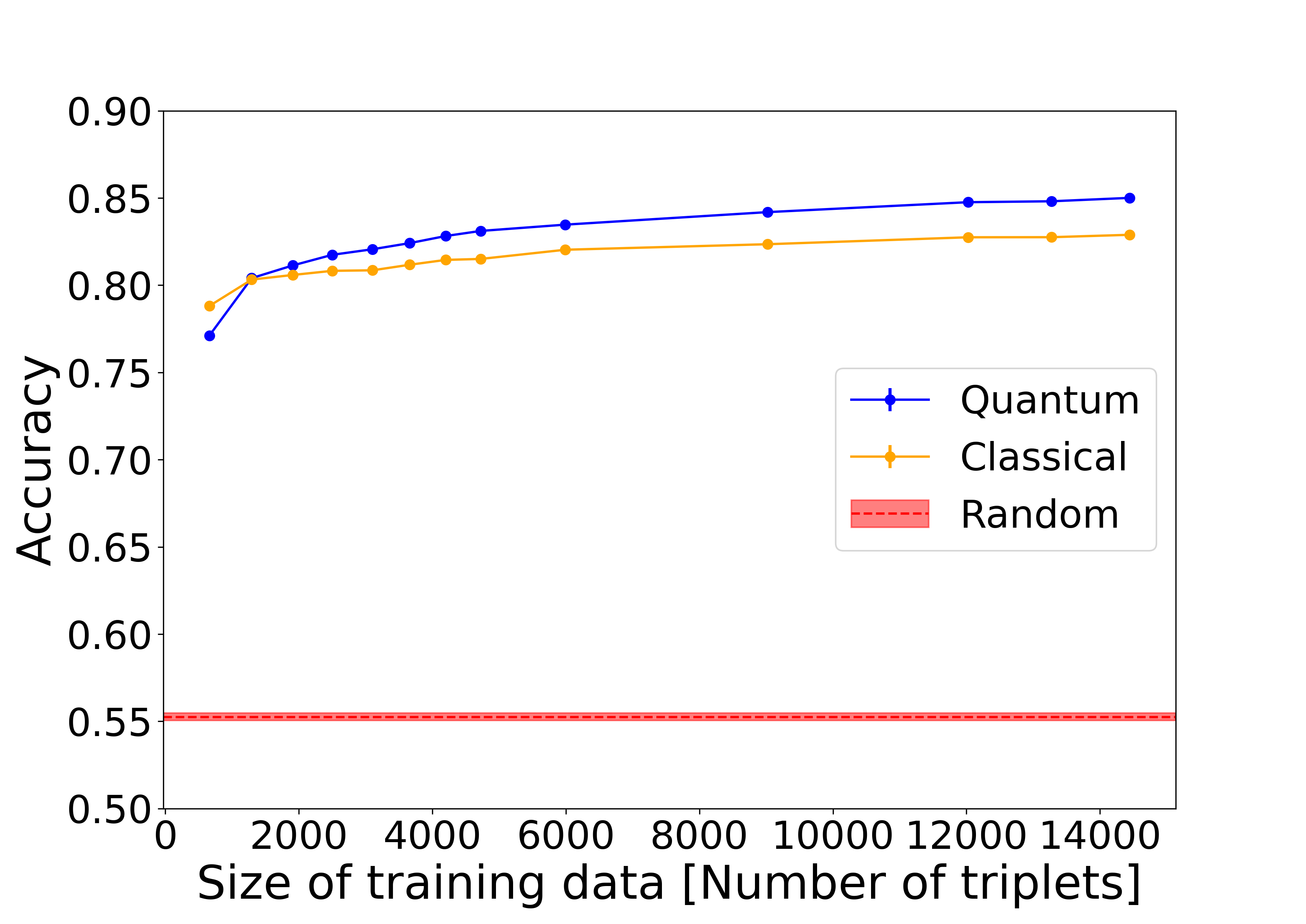}
    \caption{Accuracy to identify triplets in the inner detector barrel region as a function of the size of the training data for the quantum-estimated kernel, classical kernel and selecting random triplets from the preprocessed data. }
    \label{fig:acc_inner}
\end{figure}

This section presents results when restricting the classification of triplets to those in the innermost layers of the tracking detector, with the three hits of a triplet in the first three layers. The reduction in the total number of triplets per event allows more events to be processed before reaching the computational limit. The dataset is split into 240 events for training and 60 events for testing, equivalent to about 230,000 and 60,000 triplets, respectively. 
The efficiency and purity as a function of triplet parameters are shown in Fig.~\ref{fig:eff_pur_inner} and the accuracy is shown as a function of the size of training data in Fig.~\ref{fig:acc_inner}. 
The accuracy indicates a clearer separation between the quantum and classical performances. We see a continued trend of better purity for the quantum-enhanced classifier and a more comparable performance in terms of efficiency. 

\section{Potentials of quantum speedup}
\label{sec:q_speedup}
In general, we can assume that the evaluation of the matrix elements dominates the complexity of the SVM \cite{10.5555/1296233}. Thus, the complexity of the algorithm is $\mathcal{C} = \mathcal{O}(\beta N^2)$, where the $\beta$ factor depends on the kernel type and method used. For a single value of the kernel, the quantum complexity is $\beta_Q = \mathcal{O}(\epsilon^{-2})$ for some additive error $\epsilon$~\cite{Havl_ek_2019,Gentinetta:2022fia}. The current best classical algorithm proposed in \cite{Bravyi:2021ghk} has $\beta_C = \mathcal{O}(\epsilon^{-\frac{2}{3}}2^{\frac{2M}{3}})$. Demanding that $\beta_Q < \beta_C$ for $\epsilon = 10^{-3}$ results in a possible advantage for $M \approx 20$. 

In \cite{Gentinetta:2022fia} it has been shown that in order for the full kernel matrix to approximate the ideal kernel, the propagation of the desired accuracy into the classification with an SVM causes $\beta_Q$ to pick up a non-negligible scaling with training size $N$; $\beta_Q \rightarrow \beta_Q' = \mathcal{O}(N^{\frac{8}{3}}\epsilon^{-2})$. It is possible that a similar analysis could introduce an $N$-scaling to $\beta_C$. Regardless, it appears that at the current stage, quantum kernel estimation could provide possible advantage for small datasets where the data has many features. For the triplet classification data presented in this paper, nine features were used. We envisage that possible speedup may be obtained if we extend the length of the track segments considered in the classification task.

Another point to consider for quantum kernel estimation is the noise on current quantum devices. In \cite{Thanasilp:2022agd}, the authors show that the presence of noise can cause kernel entries produced with Eq. \ref{eq:havlicek_map} evaluated over different input data to concentrate around some fixed value. The difference between any kernel entry and that value becomes exponentially small with the number of qubits. This results in an exponential number of shots necessary to resolve kernel entries for successful training. This dependence would have to be added into $\beta_Q$ in order for the required precision to be obtained.

Some proposals for different quantum kernel estimation methods can be found in \cite{Gentinetta:2022fia}, where a probabilistic algorithm calculates only a subset of the kernel entries and \cite{Haug:2021kom} where the estimation of the entire kernel matrix scales linearly with training size $N$. 
Further studies could include empirical tests of classical and quantum complexities, study of noise effects in simulations and real quantum devices as well as implementation of other proposed quantum kernel estimation techniques in the context of high energy physics.

\section{Summary}
Reconstructing the trajectories of charged particles at particle colliders like the Large Hadron Collider is a challenging, computationally intensive problem. This is expected to become increasingly complex with the upgraded Large Hadron Collider (HL-LHC) where $\mathcal{O}(10^5)$ hits in the tracking detector must be quickly and accurately connected to form around 10,000 tracks. 
This paper presents a hybrid quantum-classical algorithm with a support vector machine (SVM) using a quantum-estimated kernel to classify track segments or seeds for this challenging track reconstruction problem. Using a publicly available dataset that simulates a generic particle detector for the HL-LHC, we apply a selection criteria to select doublets (set of two consecutive hits) and subsequently triplets (set of three consecutive hits). The proposed algorithm classifies these triplets as either belonging to a particle track or not. A comparison is made between the performance of a quantum-estimated kernel, a classical kernel, and randomly selected triplets from the dataset. A similar level of performance is achieved for the quantum and classical algorithms. However, when only the triplets from the inner part of the tracking detector are considered, the quantum algorithm shows an improvement in accuracy scores against the classical algorithm. This is promising as the innermost layers are expected to be the most important for the seeding procedure at the HL-LHC. This is the first implementation of a quantum-kernel SVM approach to the track reconstruction problem.

\section{Acknowledgements}
Sarah Malik and Tim Scanlon are funded by grants from the Royal Society. S\'{e}bastien Rettie acknowledges support from the Banting Postdoctoral Fellowship program, and the Natural Sciences and Engineering Research Council of Canada.
We also acknowledge funding from the STFC. 
We thank Simeon Hatzopoulos for helpful research assistance in visualising the TrackML dataset and Mohammad Hassanshahi for valuable discussions. 

\bibliography{QuantumTracking} 
\end{document}